\documentclass[11pt,twoside]{article}
\usepackage{asp2014}

\aspSuppressVolSlug
\resetcounters

\begin{document}

\title{Image Processing in Python With Montage}

\author{John~Good$^1$ and G. Bruce Berriman$^2$}
  \affil{$^1$Caltech/IPAC-NExScI, Pasadena, CA 91125, USA; \email{jcg@ipac.caltech.edu}}
  \affil{$^2$Caltech/IPAC-NExScI, Pasadena, CA 91125, USA}

\paperauthor{John~Good}{jcg@ipac.caltech.edu}{ORCID}{Caltech/IPAC-NExScI}{Author1 Department}{City}{State/Province}{Postal Code}{Country}
\paperauthor{G. Bruce Berriman}{gbb@ipac.caltech.edu}{0000-0001-8388-534X}{Caltech}{IPAC-NEXScI}{Pasadena}{CA}{91125}{USA}

  
\begin{abstract}

The Montage image mosaic engine has found wide applicability in astronomy research, integration into processing environments, and is an examplar application for the development of advanced cyber-infrastructure. It is written in C to provide performance and portability.  Linking C/C++ libraries to the Python kernel at run time as binary extensions allows them to run under Python at compiled speeds and enables users to take advantage of all the functionality in Python.  We have built Python binary extensions of the 59 ANSI-C modules that make up version 5 of the Montage toolkit. This has involved a  turning the code into a C library, with driver code fully separated to reproduce the calling sequence of the command-line tools; and then adding Python and C linkage code with the Cython library, which acts as a bridge between general C libraries and the Python interface. 

We will demonstrate how to use these Python binary extensions to perform image processing, including reprojecting and resampling images, rectifying background emission to a common level, creation of image mosaics that preserve the calibration and astrometric fidelity of the input images, creating visualizations with an adaptive stretch algorithm, processing HEALPix images, and analyzing and managing image metadata.
  
\end{abstract}

\section{Introduction}

Montage is an image mosaic engine that creates mosiacs an from input set of FITS images (\url{http://montage.ipac.caltech.edu}).  It is deployed as a toolkit, where each component performs one task in the creation of a mosaic, with utilities for managing and organizing files, and analyzing image metadata.  Montage was first released in 2002, and has since found wide applicability in astronony and information technology.  It has found applicability in areas such as Near Earth Object (NEO) detection, instrument performance, observation planning for missions such as JWST and NeoCAM,  and the creation of products for Citizen Science and "Big Data" Machine Learning projects. The IT community has used Montage as an exemplar application in the development and optimization of cyber-infrastructure systems, such as workfow managers and task schedulers (\citet{2017PASP..129e8006B}, \citet{2018SPIE10707E..07B}). To date in 2018, there have been 50 citations to Montage in the peer-reviewed literature and 120 citations in the IT literature.

Montage is written in ANSI-C for performance and portability. This paper describes the deployment of Python binary extensions of the Montage components. These extensions introduces the performance of Montage at compiled speeds into the flexibilty of the Python environment.

 \section{Montage and Python}

We created Python  binary extensions of 38 of the modules in version 5 of Montage. The process involved transforming the code into a C library, with driver code fully separated to reproduce the calling sequence of the command-line tools; and then adding Python and C linkage code with the Cython library, which acts as a bridge between general C libraries and the Python interface.  A uniform build across Linux platforms was achieved by compiling with a Docker container built for CentOS 5.11, which ensures consistent use of system-level functionality across all flavors of Linux. 

The extensions have been packaged as "MontagePy" and it forms part of Version 6 of Montage, released on November 12, 2018.  The package is self-contained: no additional Python tools are required to use it, although standard Python packages are valuable in examining results. A consequence of the development is that Montage can now be used as a toolkit and as a library in C, as well as under Python.

MontagePy has been developed for Python 2 and 3, but has been most thoroughly tested under Python 3.6. The package has been delivered  to the Python Package Index (PyPI) repository, and can therefore be installed with the pip package manager via the command ``{\texttt pip install MontagePy}''. The package is available on-line as  .whl files for Mac OS X and Linux distributions at \url{http://montage.ipac.caltech.edu/docs}/montagePy-UG.html for those wishing to install it directly. The C code is available in GitHub at \url{https://github.com/Caltech-IPAC/Montage}. Montage is freeely available and is attached with a BSD 3-clause license.

To assist users in processing images with Montage in Python, we have delivered a set of Jupyter notebooks that show how to use each component, and compares usage in Python with that in C.  The Jupyter notebooks are available for download at \url{https://github.com/Caltech-IPAC/MontageNotebooks}, and they can be viewed without downloading at \url{http://montage.ipac.caltech.edu/MontageNotebooks}.  

\section{Using Montage to Build A Mosaic In Python} 
The notebooks referenced above contain an example of end-to-end processing of FITS images to create a 1$^\circ$ x 1$^\circ$  mosaic of M17 in the 2MASS J-band.  This section summarizes the steps in that notebook, which includes creating the geometry of the mosaic image, reprojecting the raw images to the required output projection, modeling the sky backgrounds and correcting the images for them.  It assumes that directories have been set up to hold the raw data, the reprojected images, background differences between images, images with background corrections applied, and a working directory for the mosaicked images. 

\begin{itemize}
\item Define the parameters of a 1$^\circ$ x 1$^\circ$  output mosaic of M17 the 2MASS J band, as well as working directory for the data: {\texttt {location  = 'M 17',  size      = 1.0,  and dataset   = `2MASS J'} }
\item  Create the "template header,"  a text file that contains the WCS parameters specifying the geometry of the mosaic on the sky; file this will be written to the FITS header of the mosaic. \newline
{\texttt {rtn = mHdr(location, size, size, `region.hdr')}}
\item Download the input data for the mosaic and put them in the "raw". These can equally well already be on a local drive or be acquired through other on-line services. \newline
{\texttt {rtn = mArchiveDownload(dataset, location, size, `raw')}}
\item Create a table of metadata of the input files in the "raw" directory  \newline
{\texttt{rtn = mImgtbl(`raw', `rimages.tbl')}}
\item Reproject the input files to the required output specification \newline
{\texttt{rtn = mProjExec(`raw', `rimages.tbl', `region.hdr', projdir='projected', \linebreak 
quickMode=True)}}
\item The sky background varies between images  and must be rectified as far as is possible to a common level across all images.  Montage computes the minimum adjustments needed to make to the individual image backgrounds to bring them all in line with each other; usually this is an offset with a slope. There are several steps to applying the corrections, itemized separately below\newline

\begin{itemize}
\item Create a metadata tablefo for the reprojected images \newline
{\texttt{rtn=mImgtbl(`projected', "pimages.tbl')}}
\item  Use this metadata table to calculate the overlap area between images \newline
{\texttt{rtn = mOverlaps(`pimages.tbl', `diffs.tbl')}}
\item  Calculate the differences between images and perform fits to them \newline
{\texttt{rtn = mDiffFitExec(`projected', `diffs.tbl', `region.hdr', `diffs", `fits.tbl'}}
\item Use the fitted images to generate the background model,  a global mimimum difference which results in a set of corrections to each image: \newline
{\texttt{rtn = mBgModel(`pimages.tbl', `fits.tbl', `corrections.tbl')}}
\item Apply the background corrections to each image\newline
{\texttt{rtn = mBgExec(`projected', `pimages.tbl', `corrections.tbl', `corrected')}}
\item Create a metadata table of the background corrected images \newline
{\texttt{rtn = mImgtbl(`corrected', `cimages.tbl')}}
\end{itemize}

\item Co-add the reprojected, backgorund corrected images to create the final mosaic \newline
{\texttt{rtn = mAdd(`corrected', `cimages.tbl', `region.hdr', `mosaic.fits' )}}
\item  Make a .pngrepresentation of the mosaic for visualization: \newline
{\texttt{rtn = mViewer(`-ct 1 -gray mosaic.fits -2s max gaussian-log -out mosaic.png', "", mode=2)}}
\end {itemize}

The mosaic created by this process is shown in Figure \ref{Fig 1}.
\articlefigure[width=0.45\textwidth]{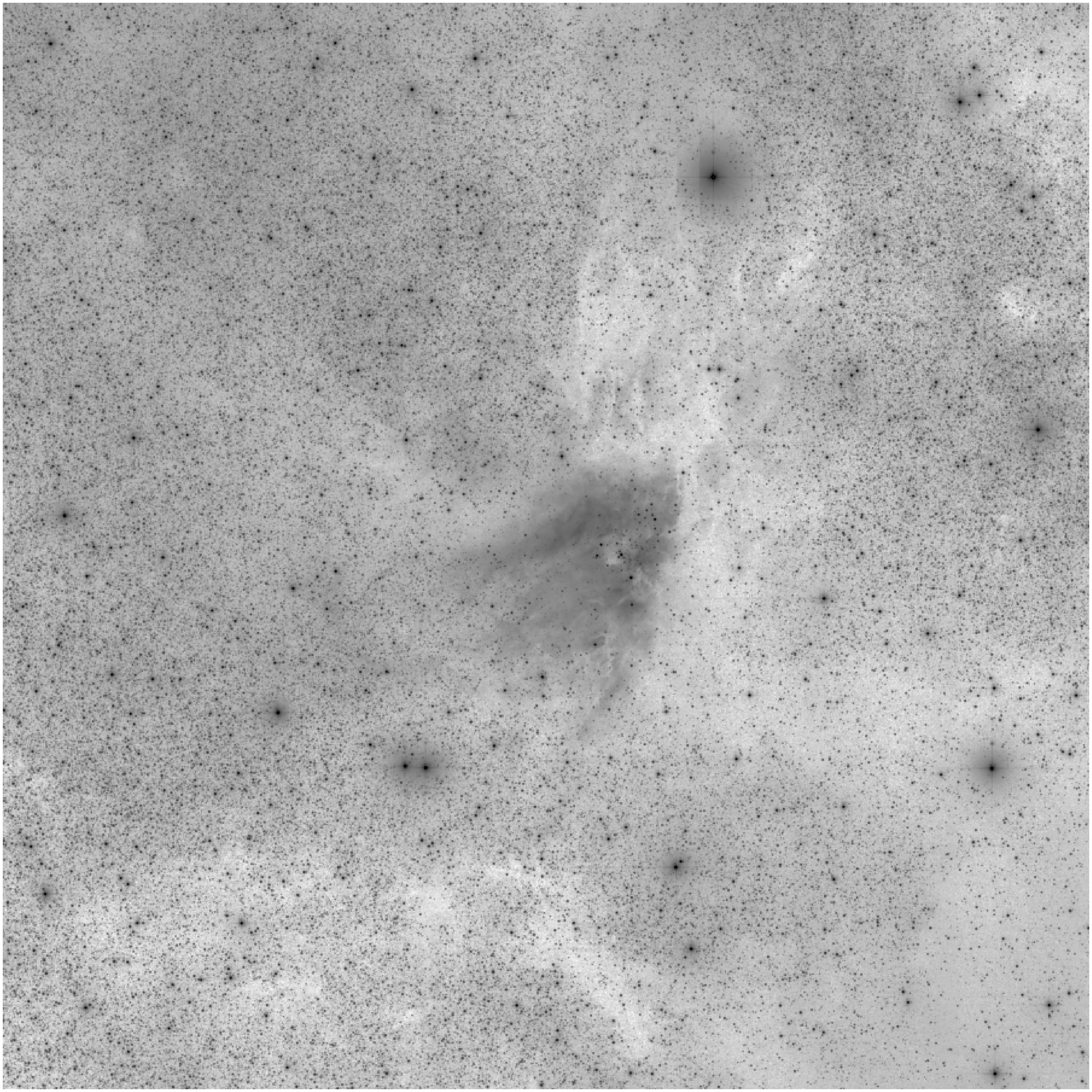}{Fig 1}{A  1$^\circ$ x 1$^\circ$  background corrected mosaic of M17 in the 2MASS J-band, created with the MontagePy package in Python 3.6}

The \texttt{mViewer} tool referenced above implements an innovative adaptive histogram equalization algorithm, which, among other things, optimizes the definition of faint structures and mid-brightness level structure \citep{2017PASP..129e8006B}. As well as gray scale ``png'' representations such as those in Figure \ref{Fig 1}, it can generate multi-color images, and can generate complex sky coverage maps, as in \cite{2018AJ....156..234C}.
\acknowledgements 
Montage is funded by the National Science Foundation under Grant Numbers ACI-1440620  and ACI-1642453, and was previously funded by the National Aeronautics and Space Administration's Earth Science Technology Office, Computation Technologies Project, under Cooperative Agreement Number NCC5-626 between NASA and the California Institute of Technology.


\bibliography{F4_v2}

\end{document}